\documentclass[twocolumn,pra]{revtex4}
\usepackage{graphicx}
\usepackage{amsmath,amssymb}
\usepackage{epsfig}
\usepackage{latexsym}
\usepackage{bm}
\usepackage{float}
\usepackage{subfigure}
\usepackage{amsfonts}

\newcommand\be{\begin{equation}}
\newcommand\bea{\begin{eqnarray}}
\newcommand\ee{\end{equation}}
\newcommand\eea{\end{eqnarray}}

\newcommand{\bdm}{\begin{displaymath}}
\newcommand{\edm}{\end{displaymath}}
\newcommand{\nn}{\nonumber \\}
\newcommand{\f}[2]{\frac{#1}{#2}}
\newcommand{\bref}[1]{(\ref{#1})}
\newcommand\h{\frac{1}{2}}

\begin{document}

\title{Hawking Evaporation is Inconsistent with a Classical Event Horizon at $r=2M$}
\author{Borun D. Chowdhury$^1$ and Lawrence M.  Krauss$^{1,2}$}
\affiliation{$^1$Department of Physics, Arizona State University, Tempe, AZ 85287}
\affiliation{$^2$School of Earth and Space Exploration, Arizona State University, Tempe, AZ 85287, USA, and Mount Stromlo Observatory, Research School of Astronomy and Astrophysics, Australian National University, Weston, ACT, Australia, 2611}
\date{\today}

\begin{abstract}

A simple classical consideration of black hole formation and evaporation times focusing solely on the frame of an observer at infinity demonstrates that an infall cutoff outside the event horizon of a black hole must be imposed in order for the formation time of a black hole event horizon to not exceed its evaporation time.  We explore this paradox quantitatively and examine possible cutoff scales and their relation to the Planck scale.   Our analysis suggests several different possibilities, none of which can be resolved classically and all of which require new physics associated with even large black holes and macroscopic event horizons:(1) an event horizon never forms, for example due to radiation during collapse (resolving the information loss problem), (2) quantum effects may affect space-time near an event horizon in ways which alter infall as well as black hole evaporation itself.
\end{abstract}


\maketitle

\section{Introduction}

Evaporating black holes present a number of paradoxes that have motivated a great deal of work in classical and quantum gravity over the past 30 years.  Most notable, as pointed out by Hawking, black hole radiance appears to be in conflict with unitarity, as pure states appear to evolve into mixed states, implying an information loss paradox that has yet to be fully resolved.  Several ideas have been proposed, from the possibility that all the information stored in a black hole is accessible at its horizon~\cite{'tHooft:1990fr,Susskind:1993if}, to the possibility that black hole event horizons do not form~\cite{Mathur:2005zp,Vachaspati:2006ki, Hawking:2014tga}. 

There is, however, a classical black hole paradox which is less often recognized.  Because of the infinite redshift factor at $r=2M$, infalling objects appear to take an infinite time to cross the event horizon in the frame defined by a distant observer, whereas the same observer will determine the lifetime of the black hole against evaporation given by the standard relation $t \approx M^3$ for a black hole of mass $M$.  This implies that in the frame defined by a distant observer, a black hole would evaporate before it forms ( i.e see \cite{Vachaspati:2006ki})

Here we explore this paradox in more detail and determine conditions on infall that might allow causality to be preserved in the process of black hole evaporation.  Our argument relies purely on classical general relativity considerations and hence is not subject to the many vagaries of interpretation often associated with considerations of quantum effects and gravity. 

We stress here several important features relevant to our discussion.  

\begin{itemize}

\item{we will not focus on what a distant observer might actually measure.  We are interested here in questions of principle, not practicality.   For example, while no one would suggest there is an information loss paradox associated with burning several pieces of paper with this article printed on them, from a practical point of view it would be essentially impossible for any observer to actually reconstruct the printed pages from the ashes.  Similarly, because of redshift effects a distant observer with finite energy sensitivity will actually cease to see an in falling object well before it approaches the horizon, and well before the black hole may be observed to evaporate.  However this does not alter the fact that in the frame of the distant observer the photons emitted by the infalling object outside the horizon do not reach the observer until well after the photons from the evaporated black hole reach the observer, independent of what the distant observer actually sees.}

\item{ we make no attempt to rely upon any possible global description of a black hole encompassing both the points of view of the in falling and distant observers, because, as is well known, no such global description exists.}

\end{itemize}

After exploring the numerical details of relevant timescales, we conclude with a brief discussion of possible implications of our analysis, all which would appear to require something like quantum effects to be significant even for the horizons around large black holes, where one would think that classical GR should be sufficient to describe space-time and associated phenomena in the vicinity of the event horizon.

\section{Infall and Observation times for a test particle near the event horizon}

Consider a massive particle starting from rest at the location $r=R  > \mathcal O(M)$.  Using time and space coordinates appropriate
to those of a distant observer located at $r$, 
the inward radial motion the four velocity is given by
\be
(\f{dt}{d\tau},\f{dr}{d\tau}) = (\f{\sqrt{1-\f{2M}{R}}}{1-\f{2M}{r}}, -\sqrt{ \f{2M}{r} - \f{2M}{R}})
\ee
where $\tau$ is the proper time.  The coordinate velocity ($v^{-1} \equiv \f{dt}{dr}$)  can be integrated to yield
\be
t_{\text{infall}}  = -\sqrt{\f{R}{2M}-1}  \int_R^{r_c} dr  \f{r \sqrt{r}}{(r-2M)\sqrt{R-r}}~.
\ee
While the result can be expressed analytically in closed form, it is not particularly illuminating.  We shall later plot specific results for a variety of cases. 

For now, we observe that near the  horizon 
\be
\f{dt}{dr} \approx -\f{2M}{r-2 M}
\ee
giving 
\be
t_{\text{infall}} = -2 M \log (r-2M) + const.
\ee

This illustrates the fact that infall times as determined in the frame of a distant observer diverge as the horizon is approached.  We can cut off these divergences by considering infall times to regions arbitrarily close to the classical horizon.  

Consider for example, cutting off infall at a Schwarzchild coordinate distance of $r=2M+1$ (recall that we are using units here where $M_P =1$).  

In order to determine the time taken for a photon emitted at this point to reach the distant observer, again in the frame of the distant observer, we need to add to the infall time estimated above, the time it takes for a signal to come out to $r=R$ from $r=2M+1$
\bea
t_{\text{outgoing}} &=& \int_{2M+1}^R \f{dr}{1 - \f{2  M}{r}}  =r+2 M \log(r-2  M) |_{2M+1}^{R} \nn
&=& R-(2M+1) +2M \log(R-2M) ~.\label{outgoingTime}
\eea

To determine specific numbers we consider $R=20M$. For large $M$ one finds
\bea
t_{\text{infall}}/M &\approx& 2 \log M +112 ~, \\
t_{\text{outgoing}}/M &\approx& 2 \log M +24 ~.
\eea

\begin{figure}[htbp] 
   \centering
   \includegraphics[width=3in]{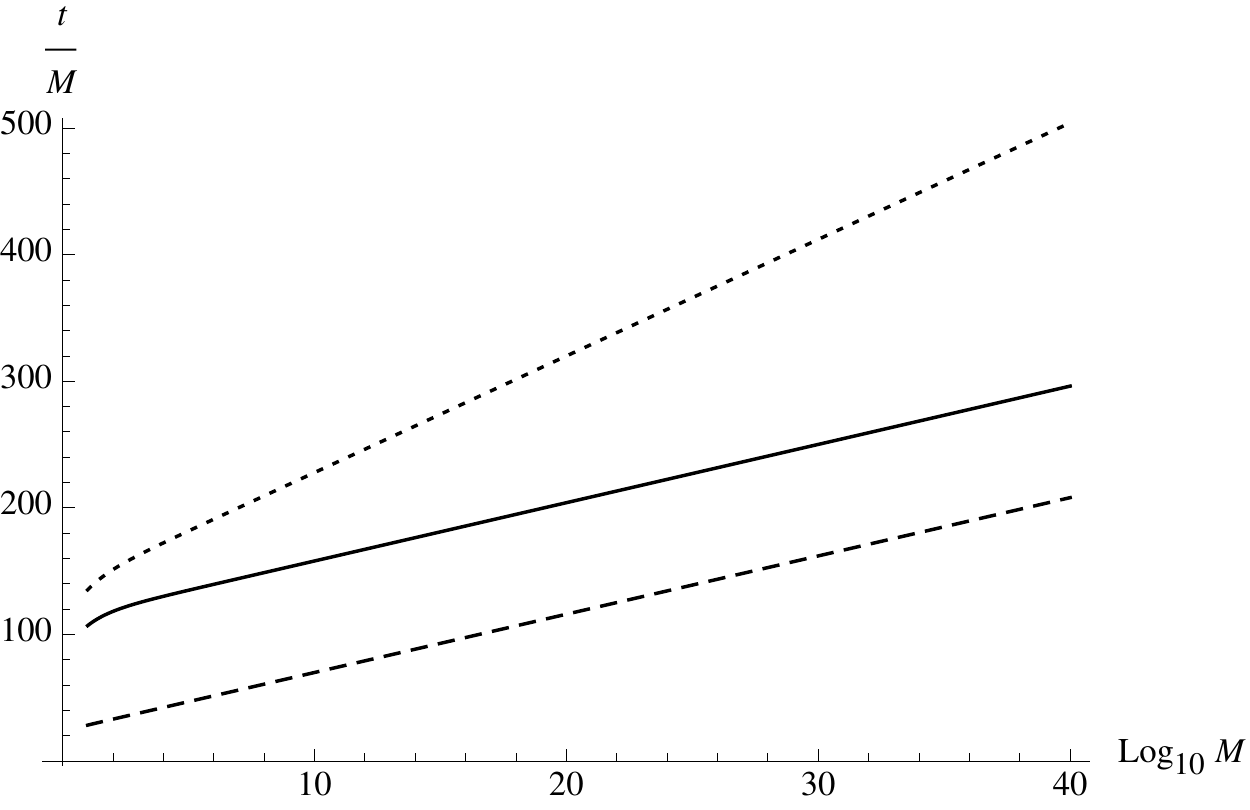} 
   \caption{The solid curve is the time to get to $r=2M+1$, starting from $R=20M$, the dashed curve is the time for a light signal to get back from that location and the dotted curve is the total time. The x-axis is logarithmic in $M$ and the times are measured in units of $M$. For reference, one solar mass is $10^{38}$.}
   \label{fig:inOutTotalTime}
\end{figure}

We plot in figure~\ref{fig:inOutTotalTime} the infall, outgoing and total time in units of mass for various masses assuming $R=20 M$. The asymptotic behaviour described above is clearly visible.

\section{Collapse time for a Self-Gravitating Spherical Shell}

For completeness, we can compare this analysis of a particle falling inside the event horizon to the more relevant case of a spherical shell of material collapsing under its own gravity, using equations of motion worked out by Israel~\cite{Israel:1966rt}. 

The speed in outside coordinates (there is a discontinuity in coordinates across the shell) is
\bea
 v^{-1} \equiv \f{dt}{dr}= - \f{\f{M}{m}-\f{m}{2r}}{1- \f{2M}{r}}  \f{1}{\sqrt{(\f{M}{m}-\f{m}{2r})^2 -1 + \f{2M}{r}}} \label{3velSelfGravShell}
\eea
where m is an integration parameter which can be taken to be the rest mass of the shell, and M is the mass parameter for the external geometry. 
 
The shell comes to a rest at
\be
R = \f{m^2}{2(m-M)}
\ee
which allows us to calculate $m$ in terms of R
\be
m=R(1 \mp \sqrt{1 -\f{2 M}{R}}) 
\ee

If we consider initial configurations of infall such that  $R \gg M$, then the two roots become
\be
m \approx M,2 R.
\ee

\begin{figure}[htbp] 
   \centering
   \includegraphics[width=3in]{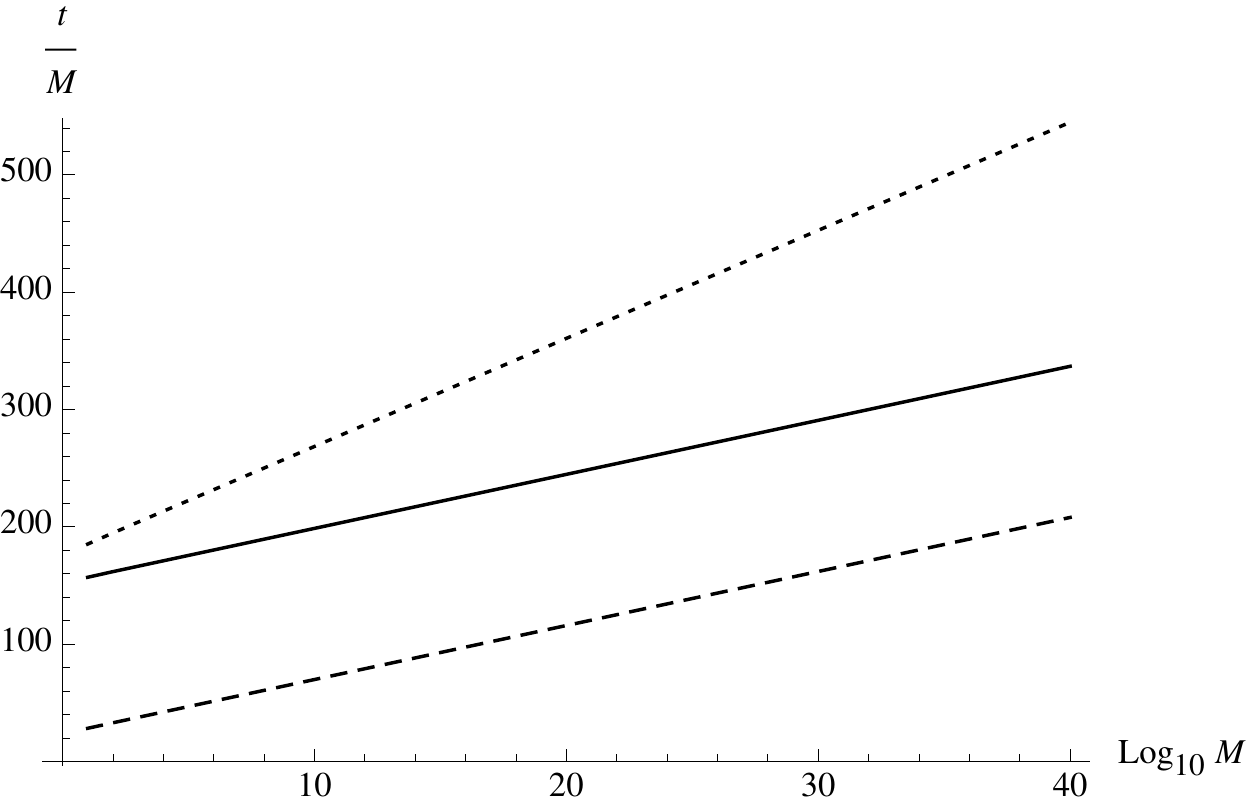} 
   \caption{The collapse time for a self-gravitating shell to reach $r=2M+1$. The convention for the curves are as given in figure 1.}
   \label{fig:inOutTotalTimeThinShell}
\end{figure}

The first root is then the appropriate one to choose in order to describe negative velocities, i.e. infall, and to get the correct ADM rest mass at infinity.   We can plug this back into~\bref{3velSelfGravShell}, and for the purposes of comparison take $r=20M$. The expression for $v^{-1}$ is not particularly illuminating, so we do not present it here.  We can again integrate this expression with respect to radial position to get the infall time, 
which we do numerically and plot in figure~\ref{fig:inOutTotalTimeThinShell}.

Comparing the two figures we see that infall times vary by at most 20-30$\%$ between the two cases, so that the analytically derived times for single test particle provide reasonable estimates to determine causality.  

\section{Temperature and Distance estimates at various Cutoff Radii}

In order to consider various cutoff distances which maintain causality in the coordinate frame of a distant observer, we consider both local temperatures and proper distances from the horizon.  
The local temperature measured by a {\em static} observer at coordinate radius $r$, outside of the event horizon, is
\be
T_r = \f{\f{1}{8 \pi M}}{\sqrt{1- \f{2M}{r}}}~. \label{localTemperature}
\ee
For $\rho \equiv r-2M \ll 2M$ this becomes
\be
T_r \approx \f{1}{2 \sqrt{2} \pi \sqrt{M \rho}}~. \label{localTemperature}
\ee
In particular at the location $r -2 M = 1$ one finds
\be
T_{2M+1}  \approx \f{1}{2 \sqrt{2} \pi \sqrt{M}}
\ee
so that the local temperature a coordinate distance of the Planck length away from the horizon is {\em well below} the Planck temperature.  

We can understand this quite simply by considering instead the proper distance from the horizon, 
\be
d_r= \int_{2M}^r \f{dr}{\sqrt{1- \f{2M}{r}}}.
\ee
 
For  $\rho \equiv r -2 M \ll 2M$ we get
\be
d_r \approx \sqrt{2M} \int_0^\rho \f{d\rho}{\sqrt{\rho}} =2  \sqrt{2} \sqrt{M \rho}~. \label{properDistance}
\ee
so a coordinate distance of $1$ from the horizon is actually $d_{2M+1} = 2 \sqrt{2M}$.

If instead we consider distance $\rho=M^{-n}$ we get
\be
d_{2 M +M^{-n}} = 2 \sqrt{2} M^{(1-n)/2}
\ee
so for $n=1$ we get the proper distance from the horizon to be Planckian.

The local temperature at this distance (i.e. for $n=1$) is
\be
T_{2M+ M^{-1}} \approx \f{1}{8 \pi M} \sqrt{2M} \f{1}{M^{-\h}} = \f{1}{2 \sqrt{2} \pi} \sim \mathcal O(1)~.
\ee
So a proper Planck distance from the horizon, corresponding to a coordinate distance $M^{-1}$, also corresponds to a local Planck temperature.

Using  \bref{outgoingTime} and we can see that the time it takes a light signal to reach a distance $\sim R$ from a distance $r-2 M \sim M^{-n}$ is, for large M
\be
t_{outgoing} \approx  2(n+1) M \log M~.
\ee
Our earlier estimates imply the infall time for massive shell will also have a similar logarithmic dependence on M.  

Since $M \log M~  \ll M^3$ it is clear that if we cut off infall at distances from the horizon comparable to regions where the local temperature is of order the Planck mass,  light emitted will reach a distant observer in significantly less time in that observer's frame that the evaporation time of the black hole as measured by that observer. 

We can ask at what distance from the horizon the outgoing time for a light ray to reach a distant observer will be of the order of evaporation timescale. That will occur when $r-2M \sim M e^{-M^2} = M e^{-S_{Bek}}$.  The physical distance corresponding to this coordinate distance is
\be
d_{2 M+ M e^{-M^2}} = 2 \sqrt{2} M e^{-M^2/2}~.
\ee

As long as we cut off infall before this distance the black hole evaporation timescale will be longer than the formation timescale. It is interesting to note that the latter factor in the distance estimate is reminiscent of a tunneling scale but    determining whether or not this is a coincidence would require a full quantum treatment. 

\smallskip
\smallskip
\smallskip

\section{Conclusions}

Our calculations explicitly demonstrate the quantitative scale of the problem associated with timescales for evaporating black holes, but of course they do not determine how to resolve this problem.   Several possibilities do suggest themselves, however.  

Perhaps the fact that the existence of a horizon at $r=2M$ implies that the evaporation time of a black hole is longer than the formation time when both are measured in the frame of a distant observer suggests a literal solution--namely that a horizon does not have time to form, as would be the case if radiation by infalling material was sufficient to cause full evaporation before horizon formation. As we have noted this would also resolve the information loss paradox associated with black hole evaporation which was the chief motivation of earlier proposals of this possibility \cite{Vachaspati:2006ki, Hawking:2014tga}.

Alternatively, some exotic quantum effects could either cause space-time fluctuations in the horizon radius, causing particles to be absorbed inside of the horizon in a finite time as observed by a distant observer.  This however would likely alter Hawking's radiation calculation, since emitted radiation at late times comes from very near the horizon, and thus would also be subject to the effects of a fluctuating horizon.   Note that in this case one might consider cutting off the Hawking radiation at late times and short distances but choosing an appropriate cutoff would require understanding the nature of the late time surface, and hence the full quantum details of late time evaporation.

Finally, perhaps some other catastrophic quantum gravity effects manifest themselves near the event horizon which would affect infall just outside of the horizon.  This possibility is reminiscent of the suggestion of fuzzballs \cite{Mathur:2005zp,Bena:2004de,Skenderis:2008qn,Chowdhury:2010ct,Bena:2013dka}, or firewalls  \cite{Almheiri:2012rt,Avery:2012tf,Bousso:2013wia,Chowdhury:2013mka,Bousso:2013ifa}.  

The first and third alternatives are actually not mutually exclusive as suggested by the following possible picture:  From the outside, infalling material appears to pile up near the pre-natal event horizon.  Processes near the horizon, which could be related to either of these possibilities, including Hawking-like pre-radiation or new quantum processes, cause this material to be observed from the outside to heat up as it collapses closer and closer to the would-be event horizon.  Ultimately an explosive, and possibly thermal burst of radiation is observed at the end of the life of the object.  In this way, no information would be lost, as nothing has lost causal contact before evaporation and an almost thermal spectrum of radiation could still emerge.

In any case, all of these possibilities imply a dramatic shift in our understanding of black hole physics and in particular the quantum processes that lead to Hawking radiation and evaporation.  While they might resolve the semiclassical temporal paradox we have focused on here, all of them beg an equally perplexing question:  Why should quantum gravity processes be relevant to understanding physics near the event horizon even for arbitrarily large black holes, where the event horizon occurs at a macroscopically large distance scale where quantum effects should naively be negligible?    

Whatever the resolution, it is interesting that classical or semiclassical considerations associated with black holes such as we have presented here point to exotic requirements for quantum gravity that may filter into even macroscopic phenomena, affecting possibly cherished classical or quantum mechanical principles.

We thank T. Vachaspati for valuable comments on the early drafts of this manuscript. This research was supported in part by a grant from the DOE.

\end{document}